\documentclass[aps,twocolumn,10pt,prc,floatfix,showpacs,preprintnumbers,amsmath,amssymb,nofootinbib,superscriptaddress]{revtex4-1}

\usepackage[normalem]{ulem}
\usepackage{wasysym}
\usepackage[dvipsnames]{xcolor}
\usepackage{graphicx}
\usepackage{subfig}
\usepackage{dcolumn}    
\usepackage{multirow, booktabs }
\usepackage{comment}
\usepackage{ulem}

\usepackage{physics, mathtools}   
\usepackage [ english ]{ babel }

\usepackage{bigstrut}
\usepackage{amsmath,amsfonts,amsthm,bm}
\usepackage{CJK}
\usepackage[pdfstartview=FitH,
            CJKbookmarks=true,
            bookmarksnumbered=true,
            bookmarksopen=true,
            colorlinks,
            linkcolor=blue,
            anchorcolor=blue,
            citecolor=blue,
            urlcolor=blue,
            ]{hyperref}

\begin{document}

\newcommand{\varR}{\mathbf{r}}
\newcommand{\varX}{\mathbf{x}}

\newcommand{\Sch}{ Schr\"{o}dinger }
\newcommand{\coeffSch}{-\frac{\hbar^2}{2m}}

\newcommand{\etal}{\textit{et al.} }
\newcommand{\redchi}{\chi^2_{red}}

\title{The complete inverse Kohn-Sham problem: from the density to the energy}

\author{A. Liardi}
\affiliation{Dipartimento di Fisica ``Aldo Pontremoli'', Universit\`a degli Studi di Milano, 20133 Milano, Italy}

\author{F. Marino}
\email{francesco.marino@unimi.it}
\affiliation{Dipartimento di Fisica ``Aldo Pontremoli'', Universit\`a degli Studi di Milano, 20133 Milano, Italy}
\affiliation{INFN,  Sezione di Milano, 20133 Milano, Italy}

\author{G.~Col\`{o}}
\affiliation{Dipartimento di Fisica ``Aldo Pontremoli'', Universit\`a degli Studi di Milano, 20133 Milano, Italy}
\affiliation{INFN,  Sezione di Milano, 20133 Milano, Italy}

\author{X. Roca-Maza}
\affiliation{Dipartimento di Fisica ``Aldo Pontremoli'', Universit\`a degli Studi di Milano, 20133 Milano, Italy}
\affiliation{INFN,  Sezione di Milano, 20133 Milano, Italy}

\author{E. Vigezzi}
\affiliation{INFN,  Sezione di Milano, 20133 Milano, Italy}

\begin{abstract}
A complete solution to the inverse problem of Kohn-Sham (KS) density functional theory is proposed. Our method consists of two steps. First, the effective KS potential is determined from the ground state density of a given system. Then, the knowledge of the potentials along a path in the space of densities is exploited in a line integration formula to determine numerically the KS energy of that system. A possible choice for the density path is proposed. A benchmark in the case of a simplified yet realistic nuclear system is shown to be successful, so that the method seems promising for future applications. 
\end{abstract}


\maketitle

\section{Introduction}


The Inverse Kohn-Sham (IKS) problem \cite{JensenWasserman,Wasserman2021} aims at reverse engineering the two steps that characterise Density Functional Theory in the Kohn-Sham direct scheme (KS-DFT) \cite{ParrYang1994,colo2020}.
In the direct DFT problem, one usually assumes an Energy Density Functional, or EDF, $E[\rho]$. Minimisation of this functional, in the form  $\delta E[\rho]=0$, provides a formulation of the ground state (g.s.) problem, while the Runge-Gross theorem  sets the formal basis for the study of excited states within a time dependent DFT framework \cite{tddft}. 

Specifically, in the KS-DFT scheme a single-particle (s.p.) representation is employed. 
Indeed, an auxiliary system of independent particles with the same g.s. density as the true interacting system is introduced. The KS density, then, is a function of the s.p. orbitals $\phi_\alpha(\varR)$ and reads
\begin{equation}\label{eq:density1}
\rho(\varR) = \sum_\alpha \phi^*_\alpha(\varR)\phi_\alpha(\varR).
\end{equation}
The EDF is customarily written as 
\begin{equation}
E[\rho] = T + F + V_{\rm ext},  
\label{eq:EDF}
\end{equation}
where $T$ is the kinetic energy of a system of independent particles, $F$ is the internal potential energy of the auxiliary KS system, and $V_{\rm ext}$ is the contribution to the energy due to an external potential $v_{ext}(\varR)$. 
A potential $v[\rho]$ is associated to $F$ by means of 
\begin{equation}
v[\rho] \equiv \frac{\delta F}{\delta \rho}. 
\end{equation} 
Then, the variational equation $\delta E[\rho]=0$, with the constraint that the orbitals form an orthonormal set, gives rise to the so-called KS equations, that are a set of one-particle Schr\"odinger-like equations. Hence, the {\it first step} of the direct KS problem consists in calculating the effective potential $v[\rho]$ from an assumed form for $F$. The {\it second step} consists in solving the KS equations in order to determine the s.p. orbitals and the corresponding density 
(\ref{eq:density1}). In short,
\begin{equation}
E[\rho] \ \ \ \rightarrow\ \ \ v[\rho]\ \ \ \rightarrow\ \ \ \rho    
\end{equation}
is a view of the direct KS problem.

The IKS problem, that is
\begin{equation}\label{eq:IKS}
\rho \ \ \ \rightarrow\ \ \ v[\rho]\ \ \ \rightarrow\ \ \ E[\rho],        
\end{equation}
is far less simple. Inverse problems are in general difficult to tackle from a mathematical and computational viewpoint.
First, we should stress that there are many more attempts and results concerning the first branch (density-to-potential, hereafter D2P) than for the second one  (potential-to-energy, hereafter denoted by P2E). 
 As for the D2P step, the first techniques to solve the IKS problem have been proposed in the context of electronic DFT in Refs. \cite{Gorling,Wang1993,vanLeeuwenBaerends,PhysRevA.50.2138}. 
This topic has recently encountered a renewed interest both in quantum chemistry \cite{KanungoBikash2019Eepf,Naito_2019,Nam2020,nam2021_code,PerezJorda2020} and in nuclear physics \cite{Accorto2020,Accorto2021}
(our group had undertaken preliminary steps before 
Ref. \cite{Accorto2020}, cf. e.g. \cite{Macocco_thesis}).
Several inversion techniques that are relevant in this context are reviewed in the useful Refs. \cite{JensenWasserman,Wasserman2021,Kumar_2019}, and inversion codes either have been  developed \textit{ex novo} \cite{nam2021_code} or have been integrated into existing libraries, such as Octopus \cite{octopus}.
Complementary to these practical implementations, 
we should also mention a recent attempt to delve more into the mathematical aspects of the IKS problem \cite{garrigue}. 

However, in spite of the indubitable technical progress \cite{KanungoBikash2019Eepf}, the first branch of the IKS problem has found \textit{per se} only a limited number of applications. 
So far, indeed, the reverse-engineered KS potentials have been mostly employed to benchmark existing models \cite{Umrigar-Gonze,Nam2020}.
In our view, the topic of utmost interest is the determination of the EDF itself. To this purpose, the P2E step is necessarily involved.
Among the few attempts in this direction, we mention that
machine learning approaches to DFT may benefit from the knowledge of the exact KS potential \cite{KanungoBikash2019Eepf}. For example, in Ref. \cite{ml_xc} it has been shown that training a neural network EDF not only on energies, but also on potentials, greatly improves its performances, at least in a simplified one-dimensional framework. 
A different line of research combines density functional perturbation theory (DFPT), where an ansatz for the EDF functional form must be assumed, and IKS to improve the initial model towards the exact EDF \cite{Naito_2019,Accorto2021}. Illustrative calculations have been performed in the case of covariant nuclear DFT \cite{Accorto2021}. 

A more direct approach to find the EDF from a solution of the whole IKS problem, as sketched in  (\ref{eq:IKS}), is our main interest in this work. The second branch, P2E, has been the subject of some theoretical investigation, although not of a full-fledged solution.
A line integration formula has been proposed by R. Van Leeuwen and E.J. Baerends in \cite{vanLeeuwenBaerendsLineintegral} as a mean to perform the functional integration 
and obtain the total energy $E$ of a given system with density $\rho$ from $v$. Such formula requires the knowledge of
the system densities not only in the ground state but also along a given path, together with the corresponding
potentials.  
The relation between the potential $v$ and the EDF has been thoroughly investigated by A.P. Gaiduk \etal in \cite{Gaiduk2009,Gaiduk2010,Gaiuduk_thesis}.  
The line integration formula has been also discussed by the authors
of Ref. \cite{Karasiev1998}, as we shall mention below.

The novel contribution of this paper consists in  outlining a possible path towards the full solution of the IKS problem, and of its implementation for a simple and yet realistic case, in nuclear physics.
In particular, we have merged the  solution of the first branch of the IKS problem (as already obtained in \cite{Accorto2020}) with the line integration formula proposed in Ref. \cite{vanLeeuwenBaerendsLineintegral}. 
We will demonstrate that, by choosing a path of conveniently scaled densities $\rho_t({\bf r})$, parametrized by a real number $t$, and by inserting densities and associated potentials in the line integration formula, we can reconstruct the total energy of the system, relative to the g.s. energy, as a function of the {\it scaling} parameter $t$  with very good accuracy. This will allow to gain some insights, as we shall discuss, about the functional form of the underlying EDF.

Our method shows some resemblance with the scheme proposed by Dobaczewski \etal in Refs. \cite{Dobaczewski_2016,Salvioni_2020}. Indeed, both approaches share the idea that information about nuclei driven outside their ground state should be used in order to probe the exact EDF. To do that, some external perturbation must be introduced. 
In Ref. \cite{Salvioni_2020} extensive \textit{ab initio} calculations have been performed, where both one- and two-body operators were added to the Hamiltonian. Our approach differs in that we generate a set of benchmark densities by means of a perturbing one-particle external potential $V_{\rm ext}(r)$, simplifying in that way the required benchmark calculations for the IKS method. 




In order to appreciate the relevance of our work, a few words about the status of nuclear DFT are in order. Nuclear DFT is a more challenging problem than electronic DFT. In the case of electronic systems, the Coulomb interaction is well-known and when building an EDF it is natural to single out the contribution from the direct Hartree term. This, together with the kinetic energy, provides the largest contribution to the EDF and leaves out only the so-called exchange-correlation part to be determined. In contrast, the effective Hamiltonian that describes the nucleon-nucleon interaction is not so accurately known (an overview of {\em ab initio} 
nuclear physics can be found in Refs. \cite{Hergert_2020,computational_nuclear,LEIDEMANN2013}). A further issue is that three nucleon forces are relevant and cannot be ignored.

In spite of all these difficulties, nuclear DFT has achieved a considerable number of successes. These are described in the recent textbook \cite{Schunck2019}. A shorter review can be found in Ref. \cite{colo2020}. Throughout the isotope chart, state-of-the-art nuclear EDFs can describe the overall trend of masses (with about 1 MeV average accuracy) and nuclear radii (with about few times 10$^{-2}$ fm average accuracy), as well as some other properties like the electric quadrupole moments or the main features of giant resonances. 
Hundreds of EDFs exist because many possible functional forms can be assumed, and many protocols to determine the EDF parameters can be designed.
It is far from obvious that new attempts to extend the EDF forms or to employ more refined fitting protocols may lead to substantial improvements \cite{unedf}.
Current EDFs are frequently based on nuclear phenomenology and suffer from large extrapolation errors \cite{ErlerJochen2012Tlot}, which are very difficult to quantify. In other words, a well established theoretical scheme for a systematic improvement does not seem to exist at present.
For those reasons, some practitioners believe that a re-thinking of the nuclear DFT strategy would be timely \cite{unedf,Furnstahl_2020,Dobaczewski_2016} (see also the work of some of us in Ref.  \cite{Marino2021}).


In this respect, non-empirical EDFs,  or even strategies to better constrain some terms of the EDFs,
would be extremely welcome. Deriving the nuclear EDF from {\em ab initio} is a possible avenue and is the subject of intense investigations \cite{Salvioni_2020,Marino2021,zurek2020}. Solving the IKS problem with experimental or {\em ab initio} densities as an input is a complementary possibility. Our work has the goal of showing that, at least in a simple situation, the IKS problem can indeed be solved. 

In Sec. \ref{sec:line_int}, the KS framework is reviewed and the line integration formula is presented. Then, in Sec. \ref{sec:remarks}, a few key aspects related to the use of this formula are discussed in detail.
Sec. \ref{sec:CV} provides a description of the method we employ to perform the D2P inversion, that is, the constrained variational (CV) method. 
In Sec. \ref{sec: results}, our method is applied to a case study. Lastly, conclusions and perspectives of our work are outlined in Sec. \ref{sec: conclusions}.


\section{The line integration formula} \label{sec:line_int}

We first remind the essential notions of DFT and clarify our notation. Useful references are \cite{ParrYang1994,colo2020,Schunck2019}.

In the KS scheme, the EDF is written as in Eq.~(\ref{eq:EDF}) as a functional of the auxiliary s.p. orbitals.
$T$ is the non-interacting kinetic energy. 


From the universal part $F$, that ultimately stems from all the inter-particle interactions,  
we define the potential $v$ (also named self-consistent potential in what follows),
\begin{equation}
\label{eq:v fdv}
v([\rho]; {\bf r}) = \frac{\delta F}{\delta\rho({\bf r})}.
\end{equation}
On the other hand, the external potential  $v_{\rm ext}({\bf r})$ and the external contribution to the total energy are related by
\begin{align}
    V_{\rm ext} = \int d^3r\ v_{\rm ext}({\bf r}) \rho({\bf r}) \\
    v_{\rm ext}({\bf r}) = \frac{\delta V_{\rm ext}}{\delta\rho({\bf r})}.
\end{align}

Applying the variational principle to Eq. \eqref{eq:EDF} leads to the following KS equations for the s.p. orbitals:
\begin{align}
\label{eq:ks eqs}
 \left( -\frac{\hbar^2}{2m}\nabla^2 + v_{\rm KS}  \right) \phi_i & = \epsilon_i \phi_i, \\
\label{eq:vks}
 v_{KS} & = v[\rho] + v_{ext}.
\end{align}
The KS potential $v_{KS}$ is the sum of the self-consistent potential and the external term. The latter is a function of the position only, while the former  also depends on the density.

After setting the framework, we can now focus on the relation between the self-consistent potential and the EDF, namely the aforementioned line integration formula.  
In Ref. \cite{vanLeeuwenBaerendsLineintegral} (cf. also 
\cite{Gaiduk2009}), it is shown that if one knows the effective potential $v[\rho]$ along a path of densities, then the 
corresponding change in the energy functional can be reconstructed. 
In particular, a one-parameter family of densities is considered. Accordingly, we shall write the densities as $\rho_t(\bf r)$, with $A \le t \le B$.
The reconstruction formula has been discussed in \cite{vanLeeuwenBaerendsLineintegral,Gaiduk2009}, which focus on electronic DFT, for the exchange-correlation part of the functional only.  
On the same grounds, we can write the following formula for $F$:
\begin{equation}\label{eq:Etot}
F[\rho_B]-F[\rho_A] = \int_A^B dt\ \int d^3r\ 
v\left([\rho_t({\bf r})], \bf r\right) \dv{\rho_t({\bf r})}{t} \ .
\end{equation}
This formula is actually holds for any functional of the density and of its gradients. We also remind that EDFs are generally written as the integral of an energy density $f$
\begin{equation}
F[\rho] = \int d^3r\ f(\bf r, \rho, \nabla\rho, \ldots) \nonumber.
\end{equation}
which is unique up to a gauge transformation, i.e. a term $\theta(\varR)$ whose volume integral vanishes.

For later convenience, we also define $I_t(R)$ by means of
\begin{equation}\label{eq:integral_fun}
    I_t(R) = \int_{0}^{R} dr r^2 \int d\Omega \,  v\left([\rho_t({\bf r})], \bf r\right) \dv{\rho_t({\bf r})}{t},
\end{equation}
where
$\Omega$ is the solid angle and $R$ is the radial upper bound for the numerical integration. We expect $I_t(R)$ to be a convergent function as $R$ goes to infinity. 
Then, \eqref{eq:Etot} can also be written as
\begin{equation}
    F[\rho_B]-F[\rho_A] = \int_A^B dt\ I_t(R\longrightarrow+\infty).
\end{equation}
In principle, any reasonable density path could be chosen in order to perform the line integral. Three specific paths were discussed in Ref. \cite{Gaiduk2009}. Our actual choice of the density path 
shall be discussed below.


The line integration formula, so far, has been mostly employed in cases where an analytical dependence of the effective potential with respect to the density is given, see \cite{Gaiduk2009}. 
The same perspective is taken by the Levy-Perdew virial relation \cite{PhysRevA.32.2010,ghosh-parr-1985} between exchange potential and exchange energy, where the energy can be deduced from the potential evaluated for just one density. In passing, we mention that line integration techniques have found applications also in constructing approximation to the non-local kinetic energy in orbital free DFT \cite{PhysRevB.100.205132,doi:10.1063/1.5023926}. 

Our idea is to move one step forward: the formula \eqref{eq:Etot} shall be applied to cases where such prior analytic knowledge  $v[\rho]$ is not available. Specifically, the relation between densities and potentials shall be defined by a numerical inversion procedure, as detailed in Sec. \ref{sec:CV}.



\section{Conceptual remarks}
\label{sec:remarks}
We present here some remarks that concern the validity of the line integration formula.


\subsection{\textit{v}-representable solutions}


The first question concerns the choice of the density path. While it is simple to design a path $\rho_t(\varR)$ that includes only $N$-representable densities, it is not possible to guarantee, in general, that the path avoids densities that are not $v$-representable \cite{KRYACHKO2014123}.
From a physical viewpoint, a reasonable and somehow conservative choice may be that of spanning a path of densities that are close to the actual ground state density of a given system or to a realistic approximation thereof, e.g., as it may be determined by an {\em ab initio} calculation.
A theorem by Kohn \cite{Kohn1983} gives theoretical support to our argument. 

As an example, in Sec. \ref{sec: results} results will be  presented for densities originating from monopole deformations of a nucleus \cite{GARG201855}. In this case, the existence of a wide literature on the effects of small variations of the shape of a nucleus around its equilibrium state can be exploited to design clever and physically motivated density paths. 
To sum up, we are confident that a sound physical intuition of the systems at hand allows to avoid the $v$-representability problem.


\subsection{Self-consistent potential}

The second question is a key issue for our discussion. The D2P solution allows us to determine the KS potential of \eqref{eq:vks}, that is,
\begin{displaymath}
    v_{KS}=v[\rho] +v_{ext}.
\end{displaymath} 
However, Eq. (\ref{eq:Etot}) specifically requires the knowledge of $v[\rho]$ alone.
While we cannot suggest a general solution, we propose a way to circumvent this difficulty in specific cases. 
Indeed, the freedom in the choice of the density path has to be exploited. 
Theoretical calculations of the systems under study, subject to a known external potential $v_{ext,t}$, can be performed and the corresponding densities $\rho_t(\varR)$ can be taken as input.
In Sec. \ref{sec: results}, we will show a demonstration of the feasibility of this approach where constrained Hartree-Fock (CHF) calculations of an atomic nucleus \cite{ring} are employed to produce the benchmark $\rho_t(\varR)$ from a given $v_{ext,t}$. In the longer term, it would be interesting to employ {\em ab initio} simulations as a benchmark. 

\subsection{Energy conservation}

In this context, an interesting identity that has been first introduced in Ref. \cite{Karasiev1998}, 
is worth discussing, because it is an useful test of our numerical procedure. 
Let us assume that $\rho_t(\varR)$ is a number-preserving path, so that all densities are normalized to the same number of particles. 
Let us consider the fundamental DFT Euler equation \cite{ParrYang1994}
\begin{equation}
\label{eq: euler dft}
    \fdv{T[\rho]}{\rho(\varR)} + v_{KS}([\rho],\varR) = \mu[\rho],
\end{equation}
where $\mu$ is the chemical potential. 
Then, we can multiply Eq. \eqref{eq: euler dft} by $\dv{\rho_t}{t}$ and integrate over both $t$ and $\varR$. Since $\mu$ is independent from $\varR$ and $\int d\varR \rho_t(\varR)$ is a constant with respect to $t$, the r.h.s. vanishes.
It follows immediately that 
\begin{equation} \label{eq:Karasiev_kinetic}
    \Delta T = - \int_{A}^{B} dt \int d\varR \, v_{KS}\left(\left[ \rho_t(\bf r)\right], \bf r \right) \dv{\rho_t(\varR)}{t}.
\end{equation}
Note that the right hand side of this equation is  different from that of Eq.~(\ref{eq:Etot}) and it is equal to $-\Delta U_{KS}= - [\Delta F+\Delta U_{ext}]$. The total variation of the energy $\Delta T+\Delta U_{KS} =0$ is conserved as it should since the chosen path comes from a minimization of Eq.~(\ref{eq:EDF}) for each value of $t$. Therefore,
this relation is a conservation law that holds on density and potential paths $\rho_t$ and $v_{KS} \left[ \rho_t \right]$. 
It is potentially well suited to test the correctness of the implementation of both the D2P step and Eq. \eqref{eq:Etot}. Indeed, as detailed in Sec. \ref{sec:CV} below, from an IKS calculation both the KS potential and the kinetic energy are obtained. One can then compare $\Delta T = T[\rho_B] - T[\rho_A]$ to the integral on the r.h.s. of Eq. \eqref{eq:Karasiev_kinetic}, thus gaining information on the relative numerical accuracy of the two procedures (Sec. \ref{sec: results}). 

\section{Constrained-variational method}\label{sec:CV}
In this Section, the formalism and the implementation of the CV method \cite{Accorto2020,JensenWasserman} are reviewed. In the KS scheme \cite{ParrYang1994}, a  system of $N$ independent particles is first considered, which is described by a set of single-particle (s.p.) orbitals $\phi_\alpha(\varR)$.
The independent-particle kinetic energy is then a functional of the orbitals and reads 
\begin{align}
    T\left[ \{\phi_\alpha(\varR)\} \right] = -\frac{\hbar^2}{2m} \sum_{\alpha=1}^{N} \int d\varR\ \phi_\alpha^*(\varR)  \nabla^2 \phi_\alpha(\varR).
\end{align}
Suppose that the ground state (g.s.) density of the system $\rho(\varR)$ \eqref{eq:density1}
be equal to a given target density $\tilde{\rho}(\varR)$. Such density may be known from experiment or an 
\textit{ab initio} calculation. 
Under these assumptions, the g.s. is characterized by the condition that the kinetic energy (as a functional of the orbitals) is minimized under the constraint $\rho(\varR) = \tilde{\rho}(\varR)$. Moreover, the additional orthonormality constraints $\int d \varR\ \phi_\alpha^*(\varR) \phi_\beta(\varR) = \delta_{\alpha \beta}$ are imposed on the orbitals. 
The problem has the form of a constrained variation of an integral  functional. The Lagrange multipliers  method allows to convert it into the free optimization of the objective functional 
\begin{eqnarray}
    &&J\left[ \{\phi_\alpha(\varR)\}; v_{KS}(\varR), \{ \epsilon_{\alpha \beta} \}\right] = T\left[ \{\phi_\alpha(\varR)\} \right] + \nonumber\\ 
    &&\hspace{3cm}+ \int d\varR\ v_{KS}(\varR) \left[ \rho(\varR) - \tilde{\rho}(\varR) \right] \nonumber\\
    &&\hspace{3cm}- \sum_{\alpha<\beta} \epsilon_{\alpha\beta} \left( \int d \varR\ \phi_\alpha^*(\varR) \phi_\beta(\varR) - \delta_{\alpha\beta} \right)\nonumber\\ .
\end{eqnarray}
We stress that $J$ is a functional of the additional variables $v_{KS}(\varR)$ and of $\epsilon_{\alpha \beta}$.
We have readily identified the  Lagrange multiplier associated to the density constraint with the KS potential $v_{KS}$. These two quantities coincide only if the minimum of $J$ has been obtained. 

By solving $\delta J = 0$ plus the subsidiary conditions, the value of the orbitals and of the multipliers can be determined. The Euler-Lagrange equations for $J$ read
\begin{align}
\label{eq: sch non canonical}
    -\frac{\hbar^2}{2m} \nabla^2 \phi_\alpha(\varR) + v_{KS}(\varR) \phi_\alpha(\varR) = \sum_\beta \epsilon_{\alpha \beta} \phi_\beta(\varR). 
\end{align}
These are s.p. \Sch equations in a non-canonical form \cite{Karasiev1998}. After these have been solved, a diagonalization of the $\epsilon$ matrix allows to determine the canonical eigenfunctions and the s.p. energies, which satisfy
\begin{equation}
\label{eq: sch}
    -\frac{\hbar^2}{2m} \nabla^2 \phi_\alpha(\varR) + v_{KS}(\varR) \phi_\alpha(\varR) = \lambda_\alpha \phi_\alpha(\varR).
\end{equation}
Thus, Eq. (\ref{eq: sch}) confirms the physical interpretation of the Lagarnge multipliers. 

In practice, works on the IKS problem have focused on spherically symmetric systems \cite{Accorto2020,Accorto2021}. As a consequence, the problem simplifies to a one-dimensional radial equation, and the labels  $\alpha,\beta$  will correspond to the usual 
quantum numbers $(nlj)$ with the standard shell-model
ordering.
The angular and spin parts of the wave functions are then known, and only the radial functions $u_\alpha(r)$ have to be determined. The orthonormality conditions have to be imposed only if $l_\alpha=l_\beta$ and $j_\alpha=j_\beta$ and read $\int_0^{+\infty}dr\ u_\alpha(r) u_\beta(r) = \delta_{n_\alpha n_\beta}$.  
We note here that the inclusion of a spin-orbit potential is left for future developments.
The KS potential is then orbital-independent. 

The numerical implementation is based on a discretization scheme, where the radial functions are evaluated on a radial mesh and the finite difference method is used to approximate the derivatives \cite{JensenWasserman}. The solution proceeds as follows.
First, the constrained optimization library IPOPT \cite{wachter_shorttutorial,Wachter2005} is used to find the orbitals $u_\alpha(r)$ that minimize the kinetic energy and  at the same time satisfy the aforementioned constraints within a given tolerance (see also Ref. \cite{Accorto2020}).
Next, we aim at determining the KS potential and the energy eigenvalues. The previously determined radial functions are plugged into the non-canonical \Sch equations \eqref{eq: sch non canonical} that now turn into a linear system, whose unknowns are $v_{KS}(r)$ on each point of the mesh and the $\epsilon_{\alpha\beta}$'s \cite{JensenWasserman}. This is an overdetermined algebraic system, with one equation for each point of the mesh and each different radial function. A least-squares algorithm is employed to solve it. Lastly, the $\epsilon$ matrix is diagonalized and the canonical solutions of \eqref{eq: sch} are thus found. The outcome of the CV method, therefore, consists in the orbitals $u_\alpha(r)$, the non-interacting kinetic energy $T$, the KS potential $v_{KS}(r)$ and the eigenvalues $\lambda_\alpha$ of the effective s.p. description of the system. 

A new code has been developed in Python which allows to perform the D2P inversion once a density is given as input. 
A technical note on the numerical inputs of the CV calculations is here in order. We have found that the convergence of the method is mostly affected by the following IPOPT parameters: the relative tolerance on the value of the objective function and the absolute tolerance on the fulfillment of the constraints \cite{wachter_shorttutorial}. The choice of the upper bound $R$ of the radial mesh is also important. 

To be physically meaningful, a radial potential should vanish as $r \longrightarrow + \infty$. It is well-known that the asymptotic behaviour of the target density directly impacts that of the corresponding potential. For example, Gaussian tails are notoriously problematic, cf. \cite{Accorto2020,KanungoBikash2019Eepf}. The densities employed in this work, however, are all well-behaved, as far as their asymptotic properties are concerned.
A further caveat relates to the fact that a potential is defined only up to a constant. Thus both $v(r)$ and $\lambda_\alpha$ must be shifted, so as to ensure that $v(r)$ goes to zero at large $r$.
Actually, the correct asymptotic behaviour of the potential is ensured by a correct reproduction of the ionization energy in atomic physics or neutron/proton separation energy in nuclear physics \cite{perdew1982,perdew1997}. This is a further check or recipe that should be applied whenever this information is available.

Lastly, we mention that some inaccuracies may occasionally show up near the border of the radial mesh. For example, the potential may take unnatural values in a limited number of points in such region. 
However, we have verified this has no impact on the energy integrals due to the exponentially decreasing behaviour of the densities.

\section{Results}
\label{sec: results}

\subsection{Physical case and numerical details}
In this Section, we will implement the strategy outlined in the previous Section 
and solve the two-step IKS problem. Namely, we will try to reconstruct
the functional starting from a set of densities $\rho_{\mu}$, with $\mu_A < \mu <\mu_B$. The densities will be 
generated by means of a set of KS calculations carried out with a Skyrme EDF, to which a harmonic external potential $v_{ext,\mu}=\mu r^2$ is added.  
Note that in the nuclear physics context these are often called self-consistent (constrained) Hartree-Fock calculations.

For each value of $\mu$, we will 
use the procedure of Sec. \ref{sec:CV}
to obtain the KS potential $v([\rho_{\mu}],r)$ by inversion (first step, D2P). We will then insert these
potentials in Eq. \eqref{eq:Etot} to reconstruct the energy difference $\Delta F = F[\rho_{\mu_B}]  - F[\rho_{\mu_A}]$
(second step, P2E). At the end, we will assess to what accuracy the procedure reproduces the benchmark 
energy difference associated with the  original Skyrme EDF. 

We highlight that the same procedure can be followed starting from a set of densities
generated by {\it ab initio} calculations, without making reference to any
pre-existing EDF. The two-step solution of the IKS problem can then lead to values of $\Delta F$ that will provide precise information on the underlying EDF.

For this first application of the method, we will present 
results obtained with  a simplified EDF. 
In particular, we benchmark our method to a $t_0-t_3$ Skyrme EDF (see e.g. Ref. \cite{Chabanat1997_part1}) with which we generate the target densities. 
This EDF is characterized by the energy density
\begin{equation}
\label{eq:total_E_t0t3}
F = \int d^3r\,\, \mathcal{F} = \int d^3r\, (\mathcal{F}_0+\mathcal{F}_1),
\end{equation}
where
\begin{gather}
\mathcal{F}_0 = \frac{1}{4}t_0 (2\rho^2-(\rho_p^2+\rho_n^2)), 		\label{eq:t0t3_first_term}\\
\mathcal{F}_1 = \frac{1}{24}t_3 \, \rho^{\alpha}(2\rho^2-(\rho_p^2+\rho_n^2)), 	\label{eq:t0t3_alpha_term}\\
\rho = \rho_p+\rho_n.
\label{eq:total_rho}
\end{gather}
Since $F$ depends on the densities, but not on their gradients, the potential takes the form 
\begin{equation}
    v_q([\rho({\bf r})],{\bf r}) = \frac{\delta F}{\delta \rho_q} = \frac{\partial \mathcal{F}}{\partial \rho_q},
\label{eq:pot_as_funct_der}
\end{equation}
where $q=n,p$. 
The EDF parameters are reported in Tab. \ref{tab:t0t3_param} for a few choices of the parameter $\alpha$. Once $\alpha$ has been chosen, the coefficients $t_0$ and $t_3$ are determined in such a way that the equation of state of symmetric nuclear matter predicted by the EDF has a minimum at the empirical saturation point, $\rho=$0.16 fm$^{-3}$ and $E=-$16 MeV. In Sec. \ref{sec: potentials} and \ref{sec: results line integ}, we test our method by employing the EDF with $\alpha=$0.32. 
To further simplify the problem, we neglect Coulomb and spin-orbit and we pick up the representative case of the $^{16}$O densities ($\rho_n=\rho_p=\rho/2$).

\begin{table}
\begin{tabular}{ 
    >{\centering}p{0.15\columnwidth} 
    >{\centering}p{0.25\columnwidth} 
    >{\centering\arraybackslash}p{0.30\columnwidth}
    }
  
    $\alpha$ & $t_0$ (MeV fm$^{3}$) & $t_3$ (MeV fm$^{3+3\alpha}$) \\ \hline
    0.16 & -3068.63 & 19578.08 \\
    0.2  & -2581.88 & 16853.70 \\
    0.32 & -1851.75 & 13124.44 \\
    1    & -1024.27 & 14602.57 \\ \hline
\end{tabular}
\caption{Parameters of the $t_0-t_3$ EDFs. }
\label{tab:t0t3_param}
\end{table}

A family of scaled densities are generated by means of constrained Hartree-Fock (CHF) calculations \cite{ring,colo_skyrme}. 
As a perturbation, a harmonic external potential $v_{ext,\mu}(r) = \mu r^2$, with $\mu$ in the range [-0.25,0.25] MeV$\cdot$ fm$^{-2}$ is employed.  The physical meaning is that of driving a scaling of the radius (and of the whole nuclear density) with respect to the unperturbed case. This approach has its original motivation in the study of monopole deformations \cite{GARG201855}. The true ground state of the system is given by $\mu=0$ and is an absolute minimum of the energy $T+F$. 

The D2P inversion is  performed by means of the CV method described in Sec. \ref{sec:CV}, which provides both the effective KS potential $v_{KS}$ and the kinetic energy associated to the neutron density.

\subsection{Potentials}
\label{sec: potentials}
\begin{figure}
    \centering
    \includegraphics[width=0.50\textwidth]{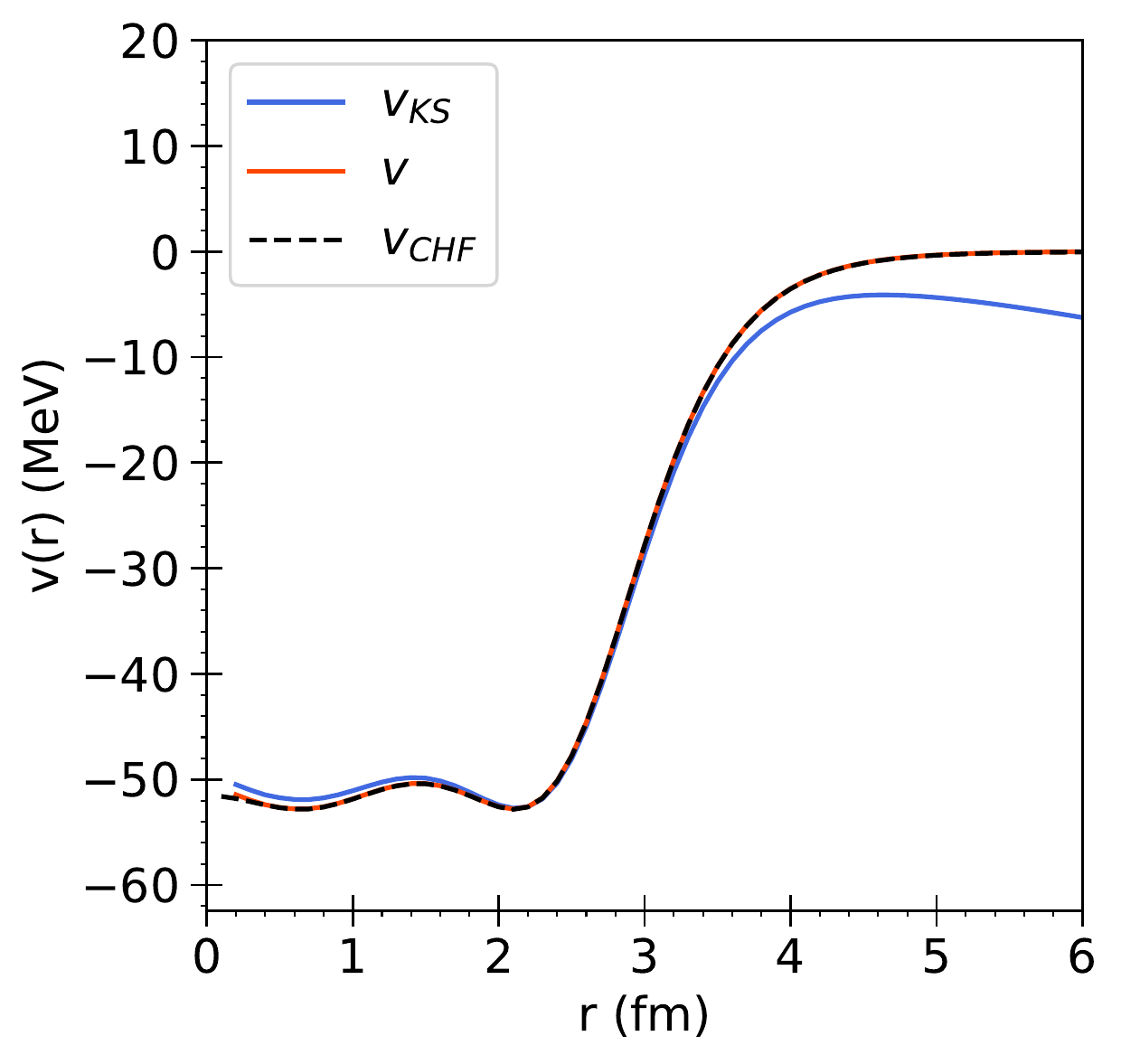}
    \caption{ The potential derived from $\rho_\mu$ by the IKS procedure ($v_{KS}$), the CHF potential $v_{CHF}$ and $v=v_{KS,\mu} - \mu r^2$ are compared for $\mu=-0.2$ MeV$\cdot$ fm$^{-2}$. 
    }
    \label{fig:pot_comparison1}
\end{figure}

\begin{figure}
    \centering
    \includegraphics[width=0.50\textwidth]{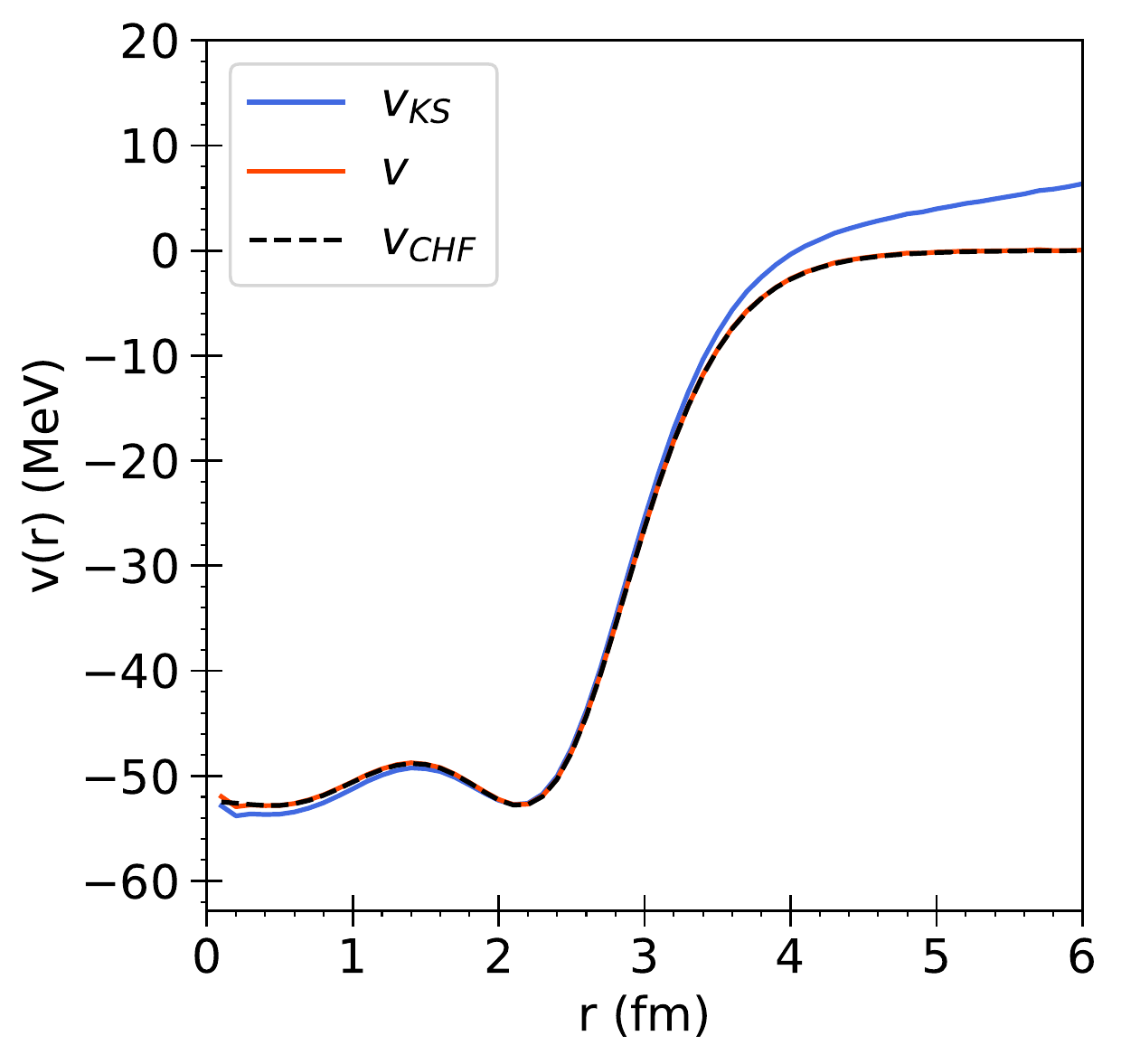}
    \caption{Same as Fig. \ref{fig:pot_comparison1}, but for   $\mu=0.2$ MeV$\cdot$ fm$^{-2}$. }
    \label{fig:pot_comparison2}
\end{figure}

To exemplify the subtle point concerning the difference between $v_{KS}$ and $v[\rho]$ (Sec. \ref{sec:remarks}), in Fig. \ref{fig:pot_comparison1} and \ref{fig:pot_comparison2} the CHF potential for either $\mu=-0.2$ or $\mu=0.2$ Mev fm$^{-2}$ is compared to the IKS potential (KS). The two functions are clearly different, while the third curve $v[\rho_\mu] = v_{KS,\mu} - v_{ext,\mu} = v_{KS,\mu} - \mu r^2$, defined by subtracting the harmonic term from the IKS potential, matches rather well the CHF potential. This proves the accuracy of the CV method.

\begin{figure}
    \centering
    \includegraphics[width=0.50\textwidth]{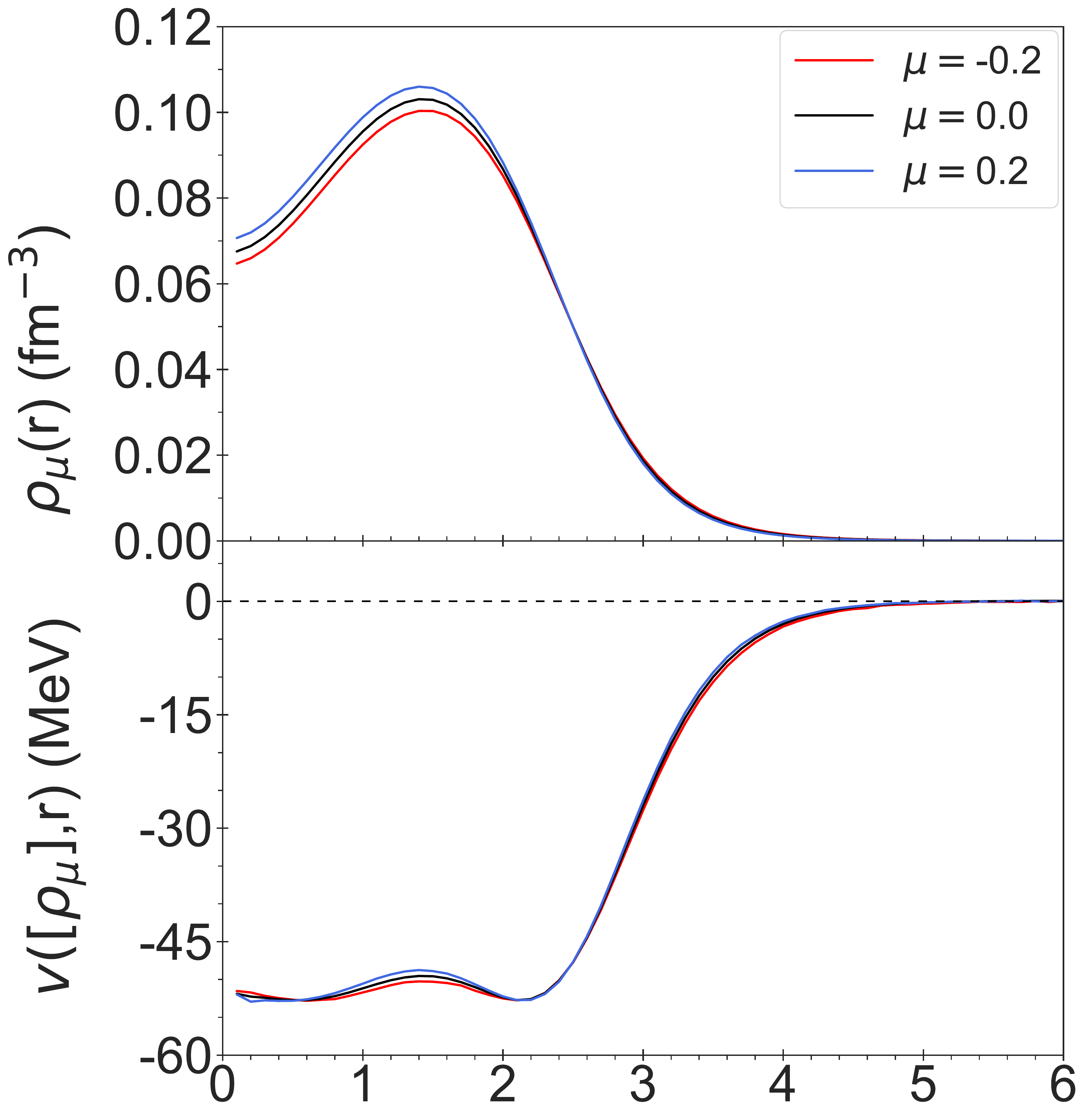}
    \caption{$^{16}$O neutron densities (top) and corresponding self-consistent potentials (bottom) for three values of the perturbation strength $\mu$ (in MeV fm$^{-2}$).  }
    \label{fig: den_pot}
\end{figure}

From now on, only the self-consistent part of the potential, $v[\rho_\mu]$, shall be displayed. 
In Fig. \ref{fig: den_pot}, three densities (top) and the corresponding potentials (bottom), determined by means of the CV inversion, are compared for three different values of $\mu$.
A positive value $\mu>0$ acts as a confining potential and as a consequence leads to a more compact density profile than in the unperturbed  case. Conversely, a repulsive external potential ($\mu<0$) leads to systems which are more spread out. Consequently, in the latter case the density peak in the interior of the nucleus is less pronounced.

\subsection{Line integration}
\label{sec: results line integ}
\begin{figure}
    \centering
    \includegraphics[width=0.50\textwidth]{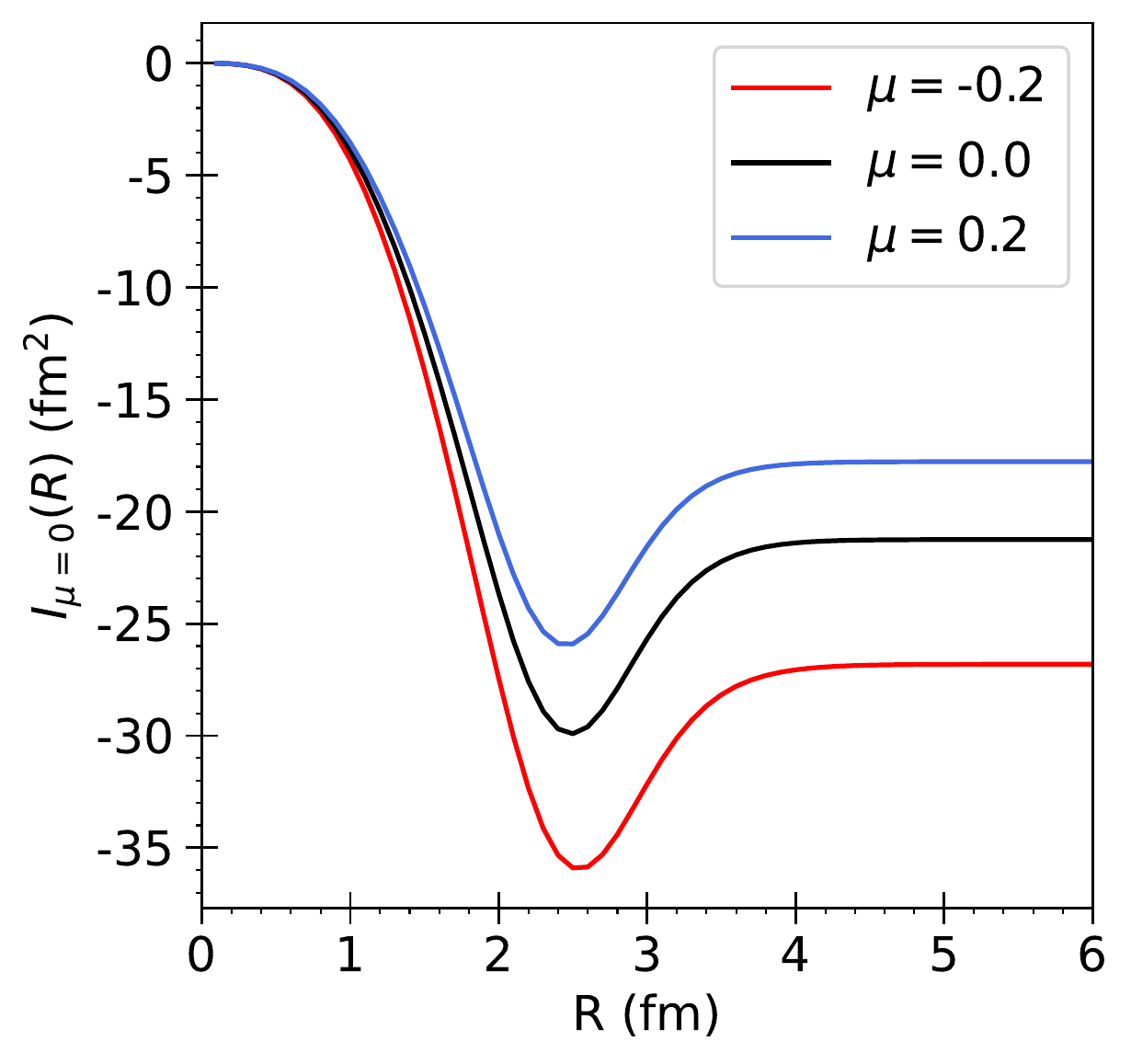}
    \caption{ $I_{\mu}(R)$ (defined in \eqref{eq:integral_fun}) as a function of the integration limit $R$ for three representitative values of $\mu$. The integral function converges asymptotically to a constant for large enough $R$.  }
    \label{fig: integrand_function}
\end{figure}

We can now move on to the study of the P2E and the line integration formula. A preliminary test is presented in Fig. \ref{fig: integrand_function}. There, the function $I_{\mu}(R)$ \eqref{eq:integral_fun} is plotted as a function of the radius $R$  in the cases $\mu=-$0.2, 0, 0.2 MeV$\cdot$ fm$^{-2}$.  Our concern here is that of verifying the asymptotic convergence  of $I_{\mu}$ to a constant for large $R$. Indeed, the convergence is quite fast and a stable result is reached already for $R=4$ fm. We remind here that ${}^{16}$O has a root mean square radius of about 3 fm and, thus, $I_{\mu}$ is expected to converge for $R$ values that scale with $A^{1/3}$. 

\begin{figure}
    \centering
    \includegraphics[width=0.50\textwidth]{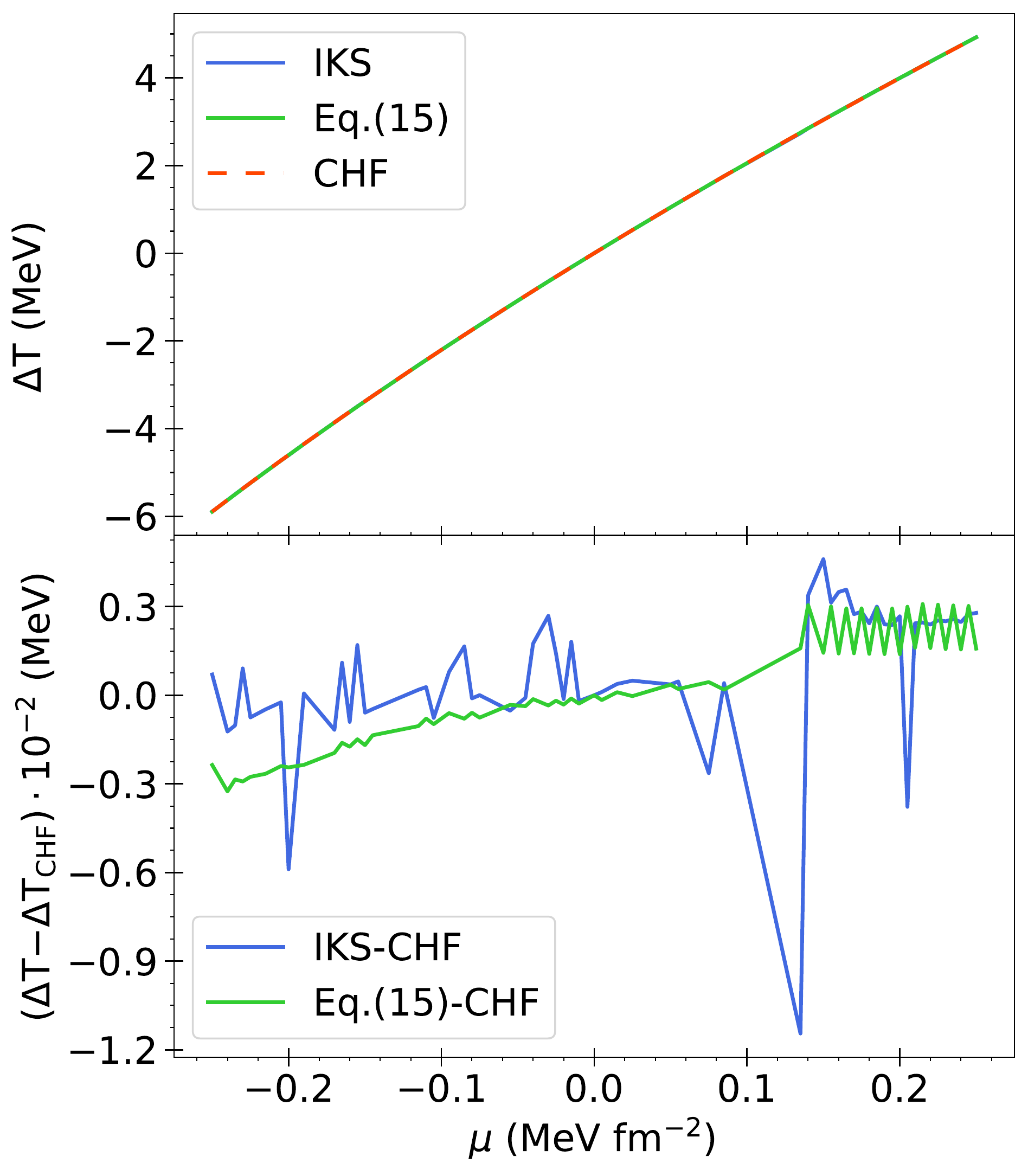}
    \caption{Top: kinetic energy differences $\Delta T(\mu) = T(\mu) - T(\mu=0)$ obtained with CHF calculations, by inversion of the densities $\rho_\mu(r)$ (IKS), and by employing Eq. \eqref{eq:Karasiev_kinetic}. Bottom: discrepancy between the CHF and both IKS and Eq. \eqref{eq:Karasiev_kinetic} results. 
    }
    \label{fig:kinetic_energies}
\end{figure}

Next, we can turn to the study of the kinetic energy and the potential energy contributions separately. 
In Fig. \ref{fig:kinetic_energies} different behaviours of the kinetic energy differences,  $\Delta T(\mu) = T(\mu) - T(\mu=0)$, are shown as functions of $\mu$. We consider two independent ways of evaluating $\Delta T(\mu)$. Indeed, one can simply subtract the kinetic energies provided by the CV method when performing inversion at $\mu=0$ and at finite $\mu$.
Otherwise, the formula by Karasiev \etal \cite{Karasiev1998}, Eq. \eqref{eq:Karasiev_kinetic}, can be exploited, where the kinetic energy is computed from the knowledge of the potential $v_{KS,\mu}$ obtained in the CV framework on a grid of intermediate values of $\mu$. 
The two estimates of the kinetic energy difference, labelled as IKS and Eq.~(\ref{eq:Karasiev_kinetic}) in the figure, are compared to the CHF results. The top panel highlights that there is a very good agreement between the three curves for all values of  $\mu$. The deviations from the CHF results are plotted in the bottom panel. It can be appreciated that the discrepancies are quite small, being of the order of 10$^{-2}$ MeV. In comparison, the neutron kinetic energy of the unperturbed system amounts to $T_{\mu=0}=148.37$ MeV (including the center of mass correction). We observe, however, that Eq. \eqref{eq:Karasiev_kinetic} guarantees an overall better accuracy. Therefore, it will be used in the following to estimate $\Delta T$. 

\begin{figure}
    \centering
    \includegraphics[width=0.50\textwidth]{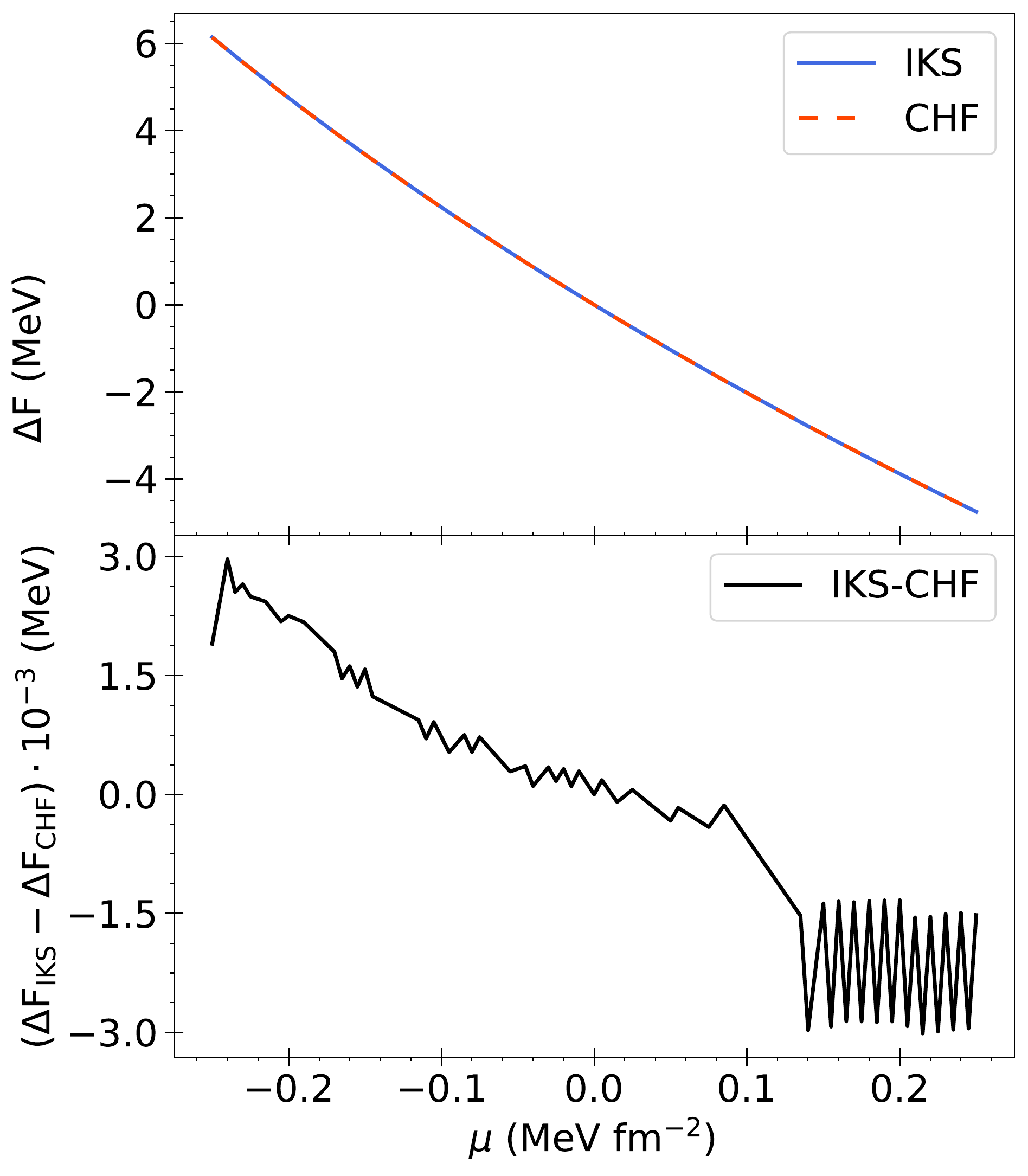}
    \caption{Top: universal energy differences $\Delta F(\mu) = F(\mu) - F(\mu=0)$ obtained with CHF calculations, by inversion of the densities $\rho_\mu(r)$ (IKS). Bottom: discrepancy between the IKS and CHF results. 
    }
    \label{fig:potential_energies}
\end{figure}

\begin{figure}
    \centering
    \includegraphics[width=0.50\textwidth]{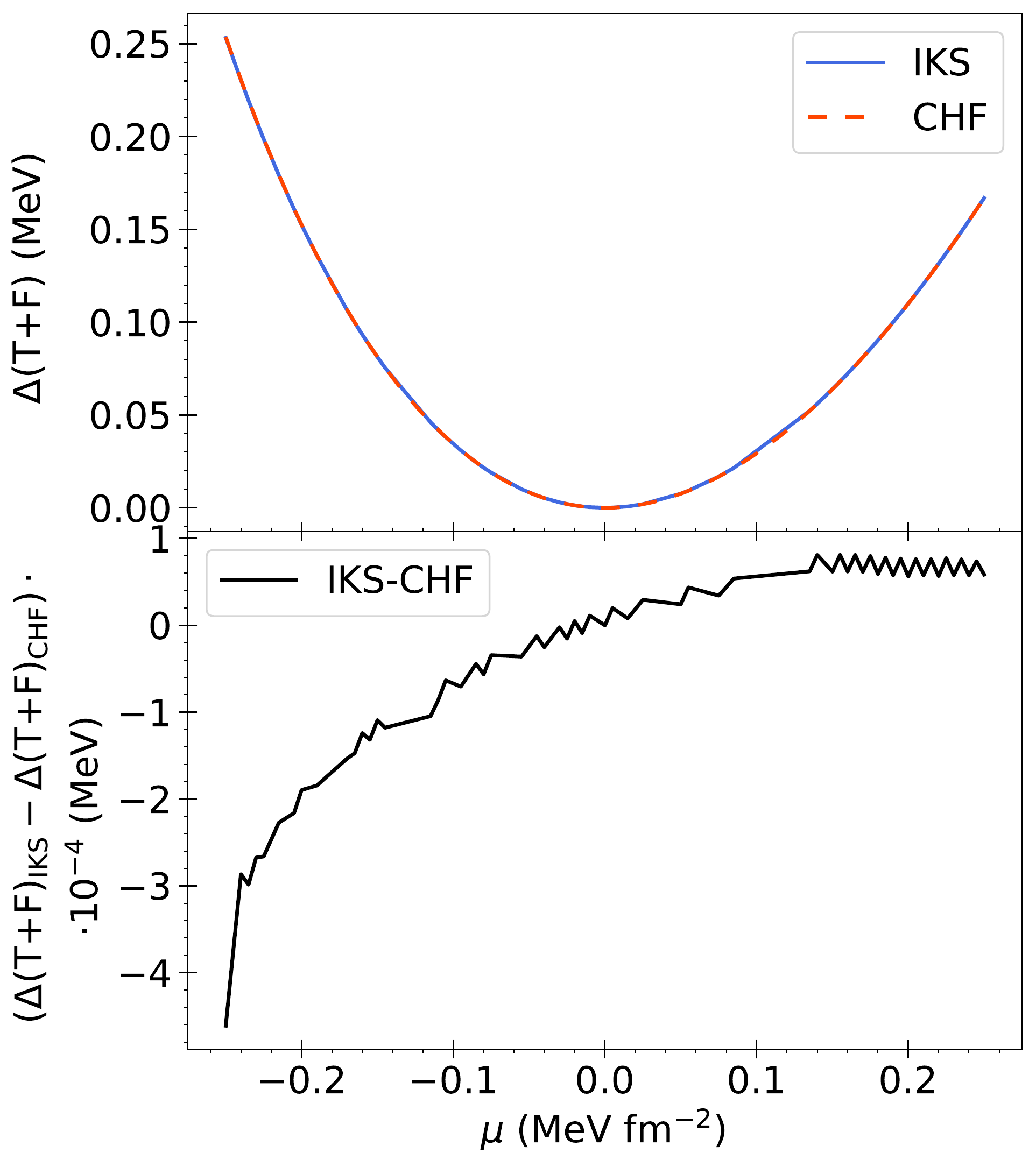}
    \caption{Top: total energy $\Delta F(\mu) + \Delta T(\mu)$ obtained with IKS and CHF calculations. Bottom: discrepancy between the IKS and CHF results. 
    }
    \label{fig:total_energies}
\end{figure}

We now compute the universal term $F$ or, more precisely, $\Delta F(\mu) = F(\mu) - F(\mu=0)$, by the line integration method introduced in Sec. \ref{sec:line_int} according to  Eq.~\eqref{eq:Etot}. Fig. \ref{fig:potential_energies} shows that the reconstructed IKS curve is in excellent agreement with  the CHF values obtained with the original Skyrme EDF, with discrepancies of about $10^{-3}$ MeV.
This is rather encouraging, since it implies that the line integration has been implemented to a high numerical precision. At the same time, it is a strong support to the reliability of the IKS machinery. 
This is confirmed by Fig. \ref{fig:total_energies}, where we display the total energy difference $\Delta F(\mu) + \Delta T(\mu)$, accurate up to a fraction of keV.

\begin{figure}
    \centering
    \includegraphics[width=0.50\textwidth]{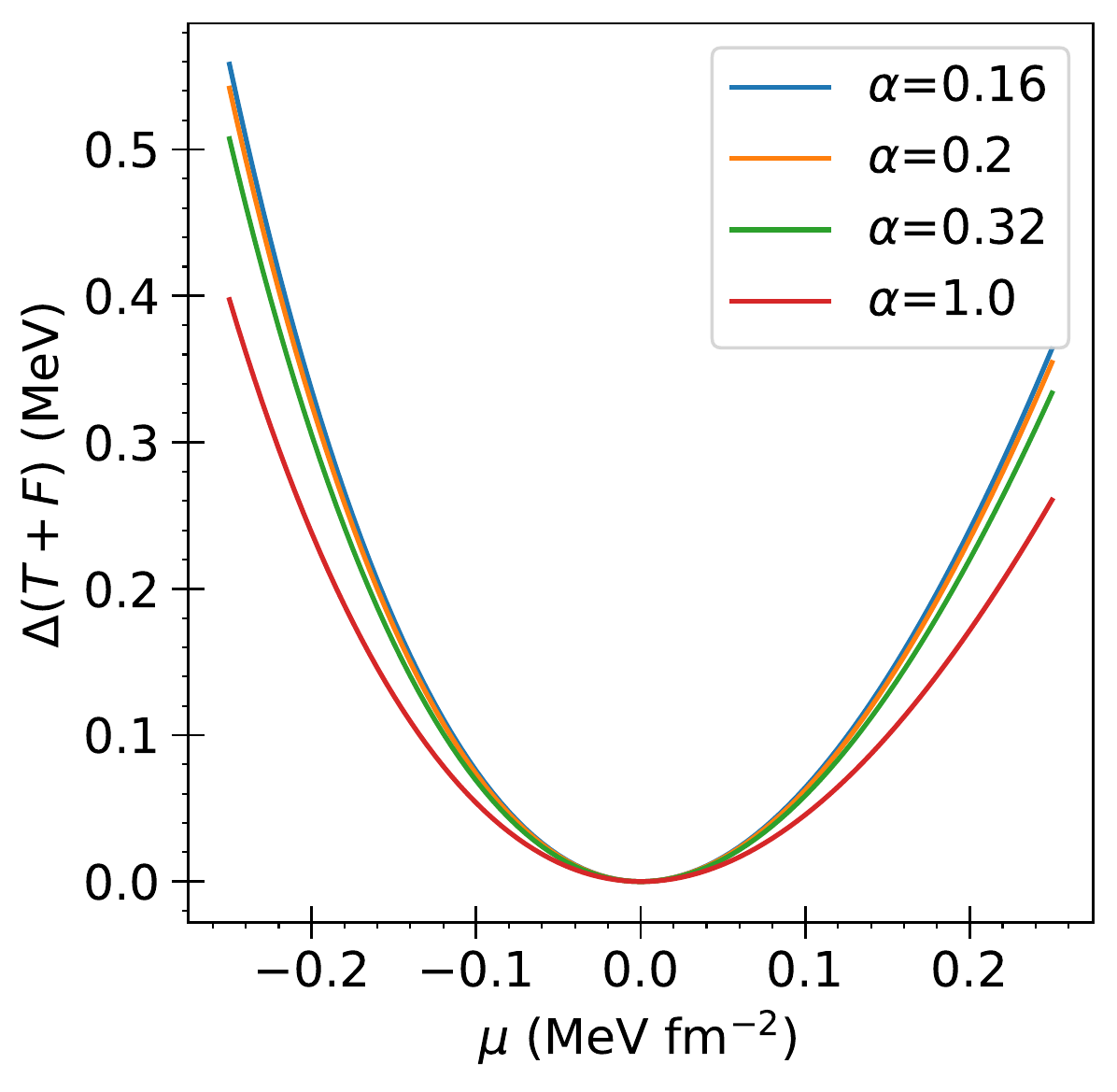}
    \caption{
    The total energy $\Delta F(\mu) + \Delta T(\mu)$ reconstructed with the two-step IKS method
    and associated with the value $\alpha=0.32$
    is compared to the values obtained with $t_0-t_3$ EDFs characterized by different values of
    $\alpha$.
    }
    \label{fig:total_en_alpha}
\end{figure}

A question now arises: how much information have we actually gained about the original
EDF, by solving the two-step inverse IKS problem? 
We will limit ourselves
to consider the family of $t_0-t_3$ Skyrme EDFs, obtained by varying the value of the 
exponent $\alpha$ (cf. Tab. \ref{tab:t0t3_param}) and by
assuming that they are determined in symmetric nuclear matter,
as discussed above  (cf. Tab. \ref{tab:t0t3_param}).
In other words, we will not discuss how to determine
the absolute strengths ($t_0$ and $t_3$) of the different terms in the underling $F[\rho]$, 
but only if we can be sensitive to the exponent of the density dependence. To this aim, we compare in Fig.~\ref{fig:total_en_alpha}
the values obtained for $\Delta T+\Delta F$ as a function  of $\mu$ for a few different values of $\alpha$.
The resolving power of the method will depend on the range of $\mu$ taken into account in the calculation. Nonetheless, the figure already shows that, for a modest change in $\mu$, our  benchmark value $\alpha=0.32$ can be rather clearly distinguished from  $\alpha$=0.16 and $\alpha$=1. This type of information may become instrumental in building new EDFs based on {\it ab inito} calculations.

\section{Conclusions and perspectives}
\label{sec: conclusions}
A complete solution to the inverse Kohn-Sham problem of Density Functional Theory has been proposed. 
First, the effective KS potential associated to a given ground state density is determined. Second, a path of perturbed densities is chosen, and the knowledge of the associated KS potentials is exploited to compute the difference between the energies of the perturbed and unperturbed states. 

Benchmark calculations have been performed in a case relevant for nuclear physics. In particular, we have shown that, for a simple $t_0-t_3$ Skyrme EDF, and perturbing the system with an external harmonic potential, this method allows to reconstruct the energies with a rather good numerical accuracy. 

These promising results open up a number of perspectives.
First, one should apply our method using {\em ab initio} calculations as an input. The feasibility of these calculations when the nuclear systems are perturbed by external potentials, and the assessment of the associated theoretical uncertainties, are issues that are mandatory to clarify.

In principle, accurate {\em ab initio} calculations of nuclear densities perturbed by a variety of external potentials contain a lot of useful information to improve existing EDFs. In practice, one could apply the IKS methodology outlined in the present work to produce a set of {\em ab initio} metadata on how energies change under the action of perturbations. 

In our work, we have not
tackled the problem of how to fully determine the EDF
from these metadata. When {\em ab initio} inputs are available, we envisage to use also the resulting absolute values of the energy. Implementing appropriate fitting algorithms on these absolute energies may allow obtaining a set of novel constraints on the different terms of the EDF. As the convergence of this procedure is not guaranteed, because the choice of the different terms is the result of an ansatz, and the EDF which is compatible with the {\em ab initio} input may not correspond to that ansatz, one
could devise as an alternative a machine learning algorithm (see, e.g., \cite{ml_xc}). This should be able to select the terms in the EDF that are more likely to appear, and quantitatively estimate their relevance.

In a later stage, further studies could be envisaged by considering a set of different external potentials that are coupled to spin densities, isospin densities, or other types of densities. This investigation could shed light on the terms of the nuclear EDF that are not
merely sensitive to the total density. 

\bibliography{bibliography.bib} 

\end{document}